\documentclass[aps,prd,notitlepage,superscriptaddress,longbibliography,nofootinbib]{revtex4-2}
\usepackage[dvipdfmx]{graphicx}
\usepackage{float}

\usepackage{dcolumn}    
\usepackage{bm}         
\usepackage{hyperref}   
\usepackage{bmpsize}
\usepackage[english]{babel}
\usepackage[skip=1pt, justification=raggedright]{caption}
\usepackage{graphicx}
\vspace{-10pt}
\usepackage{amsmath}
\usepackage{subcaption}
\hypersetup{
    colorlinks=true,
    citecolor=blue,
    linkcolor=blue,
    urlcolor=blue
}

\begin{document}

\title{Prospects for Observing Gravity-gradient Noise and Earthquake Gravity Signals with CHRONOS}

\author{Mario Juvenal S. Onglao III}
\thanks{Corresponding author: iyuki@ncu.edu.tw}
\affiliation{National Institute of Physics, University of the Philippines - Diliman, Quezon City 1101, Philippines}
\affiliation{Center for High Energy and High Field (CHiP), National Central University,  Taoyuan, Taiwan}

\author{Yuki Inoue}
\affiliation{Center for High Energy and High Field (CHiP), National Central University,  Taoyuan, Taiwan}
\affiliation{Department of Physics,National Central University, Taoyuan, Taiwan}
\affiliation{Institute of Physics, Academia Sinica, Taipei, Taiwan}
\affiliation{Institute of Particle and Nuclear Studies, High Energy Acceleration Research Organization (KEK), Tsukuba, Japan}

\author{Daiki Tanabe}
\affiliation{Center for High Energy and High Field (CHiP), National Central University,  Taoyuan, Taiwan}
\affiliation{Department of Physics,National Central University, Taoyuan, Taiwan}
\affiliation{Institute of Physics, Academia Sinica, Taipei, Taiwan}
\affiliation{Institute of Particle and Nuclear Studies, High Energy Acceleration Research Organization (KEK), Tsukuba, Japan}

\begin{abstract}
Ground-based gravitational-wave detectors operating in the sub-Hertz regime are expected to be strongly limited by environmental gravity-gradient fluctuations, commonly referred to as Newtonian Noise (NN). At the same time, this frequency band provides unique opportunities to probe terrestrial gravitational perturbations associated with seismic and atmospheric processes. In this work, we investigate the feasibility of using the proposed Cryogenic sub-Hz cROss torsion-bar detector with quantum NOn-demolition speed meter (CHRONOS) as a platform for studying gravity-gradient noise and detecting prompt gravitational signals from earthquakes.

We model gravity-gradient contributions from Rayleigh-wave-induced seismic fields, atmospheric infrasound fluctuations, and transient mass redistribution during earthquakes, and project these onto the CHRONOS torsion-bar response. CHRONOS achieves a peak strain sensitivity of order $\sim 10^{-18}\,\mathrm{Hz^{-1/2}}$ near $\sim 2\,\mathrm{Hz}$. Rayleigh-wave NN is found to be the dominant environmental contribution below approximately $0.5\,\mathrm{Hz}$, while atmospheric NN remains several orders of magnitude smaller throughout the frequency range considered.

We further assess the detectability of prompt gravitational signals from earthquakes. For a representative $M_w = 5.2$ event, sources within approximately $90$ km may produce detectable signals. At $40$ km distance, we obtain a signal-to-noise ratio of $\mathrm{SNR} \approx 3.62$ integrated over the sub-Hz band, with a corresponding strain amplitude reaching the CHRONOS sensitivity curve around $\sim 0.2$–$0.6\,\mathrm{Hz}$. The gravitational signal is expected to precede seismic P-wave arrival by several seconds, depending on the assumed propagation velocity.

These results demonstrate the potential of CHRONOS to probe both gravity-gradient noise and transient geophysical gravity signals in the sub-Hertz regime.
\end{abstract}

\maketitle

\section{Introduction}
\label{sec:introduction}
Gravitational waves (GWs), propagating perturbations of spacetime generated by accelerating massive bodies, were first directly observed in 2015 by Advanced LIGO \cite{abbott2016observation}. This discovery established gravitational-wave astronomy as a new observational probe, enabling studies of strong-field gravity, compact-object populations, and astrophysical and cosmological phenomena that are inaccessible through electromagnetic observations alone.

Second-generation ground-based detectors, including Advanced LIGO and Virgo, have now approached their design sensitivities \cite{buikema2020sensitivity,capote2025advanced}, leading to the increase in detection rates of GWs, and the construction of rich event catalogues. These detectors have been designed for sensitivity in the high frequency range of 10 Hz to 10 kHz. To probe more massive sources and to extend observational coverage, further improvements in sensitivity are required, targeting the sensitivity gap in the 0.1–10 Hz frequency range. In response, a slew of next-generation (third-generation) detectors sensitive to different frequency ranges are currently under development, such as Cosmic Explorer \cite{reitze2019cosmic} and the Einstein Telescope \cite{punturo2010einstein}, along with alternative concepts including torsion-bar antennas \cite{ando2010torsion,mcmanus2017mechanical}. Improvement in sensitivity by refining calibration techniques have also been proposed\cite{inoue2026improvingcalibrationaccuracytorque}. These efforts aim to expand the accessible bandwidth of gravitational-wave observations by improving sensitivity particularly in the sub-Hz regime.

The Cryogenic sub-Hz cROss torsion-bar detector with quantum NOn-demolition speed meter (CHRONOS) \cite{inoue2026opticaldesignsensitivityoptimization, tanabe2025torque, SPP-2026-3A-02, SPP-2026-3A-03, SPP-2026-3A-06} is a proposed instrument designed to operate in the 0.1 to 10\,Hz frequency band, with a target strain sensitivity of $h \sim 10^{-18}\,\mathrm{Hz^{-1/2}}$ near $2\,\mathrm{Hz}$. The detector concept stands on three pillar technologies: cryogenic operation to suppress thermal noise, suspended cross torsion-bar configuration to decrease resonant frequency and suppress seismic noise, and quantum non-demolition speed meter to reduce quantum back-action noise. The speed meter is composed of a Sagnac interferometer with triangular cavities implementing a balanced-homodyne readout. Together, these three technologies aim to enable high-sensitivity measurements in a frequency range that remains largely inaccessible to current detectors.

The scientific potential in this sub-Hz band is substantial. Intermediate-mass black hole binaries are predicted to emit gravitational radiation that sweeps through the 0.1--10 Hz range prior to merger, providing early inspiral information inaccessible to higher-frequency detectors. Detection of the stochastic gravitational-wave background may also benefit from improved low-frequency sensitivity, where cosmological and astrophysical backgrounds can exhibit distinct spectral features.

In addition to these astrophysical targets, detectors operating in the sub-Hz band are expected to be sensitive to terrestrial gravity perturbations generated by seismic and atmospheric mass-density fluctuations. These perturbations produce gravity-gradient signals that couple directly to the detector through gravity and therefore cannot be suppressed by conventional mechanical isolation systems.

Beyond astrophysics, sensitivity in this band offers the possibility of earthquake early warning through the detection of prompt time-varying gravity perturbations that travel at the speed of light, preceding the arrival of seismic P-waves \cite{harms2013low}. These prompt gravitational signals, often treated as a source of Newtonian Noise (NN) in gravitational-wave detectors, can instead be considered as signals of interest for gravity-based early warning.

However, a fundamental limitation in the sub-Hz regime targeted by second generation gravitational-wave detectors arises from gravity-gradient, or Newtonian, noise \cite{harms2013low}. NN is generated by fluctuations in the local gravitational field caused by mass-density perturbations in the surrounding environment, particularly those associated with Rayleigh waves or the atmosphere. Since this coupling occurs directly through gravity, it cannot be suppressed by mechanical isolation alone. As such, NN is expected to become one of the dominant sensitivity limitations below a few hertz and represents a major challenge for future low-frequency gravitational-wave detectors such as CHRONOS.

To address this limitation, numerous techniques to characterize and mitigate NN have been developed \cite{coughlin2016towards,harms2015newtonian}. Accurate modeling and mitigation of NN require improved understanding of its physical sources and site-dependent characteristics. While analytical descriptions of Rayleigh-wave coupling and atmospheric contributions exist, uncertainties in subsurface structure and spatial correlations limit predictive power. Direct gravity-gradient measurements, combined with refined modeling, are essential to quantify the true noise floor and to develop noise-correction strategies using seismic and atmospheric sensor arrays.

In this work, we investigate the feasibility of using CHRONOS as a gravity-gradient observatory in the sub-Hz regime. We model gravity-gradient signals arising from Rayleigh-wave seismic fields, atmospheric infrasound fluctuations, and transient earthquake-induced mass redistribution, and project these signals onto the detector response through a detector-specific mechanical coupling model. Our goal is twofold: to assess the impact of Newtonian noise on the projected CHRONOS sensitivity, and to evaluate the potential of CHRONOS to directly measure terrestrial gravity perturbations that are traditionally regarded as environmental noise.

\section{CHRONOS model}
\label{sec:CHRONOS}

This section describes the design of CHRONOS detector used in the sensitivity calculations. Relevant parameters are summarized in Table~\ref{tab:CHRONOSparams}, while a complete description of the detector design can be found in the CHRONOS white paper \cite{inoue2026chronosscienceprogram}.

The interferometer is based on a cross-shaped torsion-bar configuration with identical optical systems arranged on orthogonal meter-long bars. This geometry provides rejection of common-mode environmental disturbances that are particularly significant in the sub-Hz frequency band. Massive end test masses (ETMs) are attached near the ends of each bar, increasing the moment of inertia while maintaining a low torsional resonant frequency. The entire structure is suspended by sapphire fibers and behaves as a rigid torsional oscillator about the vertical axis.

For the gravity-gradient calculations presented in this work, the dominant degree of freedom is the yaw rotation angle $\theta$ about the suspension axis. External gravity-gradient fields generate differential gravitational forces on the ETMs, producing a torque that drives the torsional motion of the bar. Neglecting higher-order elastic modes, the rotational dynamics are described by

\begin{equation}
I\ddot{\theta}
+\frac{I\omega_{\rm tb}}{Q}\dot{\theta}
+I\omega_{\rm tb}^{2}\theta
=
\tau_{\rm ext},
\label{eq:torsion}
\end{equation}

where $I$ is the moment of inertia of the torsion bar, $\omega_{\rm tb}=2\pi f_{\rm tb}$ is the torsional resonant angular frequency, $Q$ is the mechanical quality factor, and $\tau_{\rm ext}$ denotes the external torque induced by gravity-gradient perturbations.

In the frequency domain, $f$,  Eq.~(\ref{eq:torsion}) can be written as

\begin{equation}
\theta(f)
=
\chi_{\theta}(f)\,
\tau_{\rm ext}(f),
\end{equation}

where the torsional susceptibility is given by

\begin{equation}
\chi_{\theta}(f)
=
\frac{1}
{I\left(
\omega_{\rm tb}^{2}
-\Omega^{2}
+i\Omega\omega_{\rm tb}/Q
\right)},
\label{eq:torsion_susceptibility}
\end{equation}
and $\Omega = 2 \pi f$.
The susceptibility in Eq.~(\ref{eq:torsion_susceptibility}) determines how external gravity-gradient perturbations are converted into angular motion and forms the basis of the detector response calculations described in the following section.

The relevant parameters used in this work are summarized in Table~\ref{tab:CHRONOSparams}.

\begin{table}
    \centering
    \begin{tabular}{c|c}

    \textbf{Parameter} & \textbf{Value} \\
    \hline
    Torsion bar length ($L$) & 1.0 [m] \\
    \hline
    End Test Mass ($m_{\rm end}$) & 171.0 [kg] \\
    \hline
    Moment of inertia ($I$) & 19.9 [kg\,m$^2$] \\
    \hline
    Torsional resonance frequency ($f_{\rm tb}$) & 0.0037 [Hz] \\
    \hline
    Torsional mode quality factor ($Q$) & $10^7$ \\
    \hline
    \end{tabular}
    \caption{CHRONOS design parameters relevant to this study.}
    \label{tab:CHRONOSparams}
\end{table}

\section{Gravity-Gradient Formalism}
\label{sec:GG}

This section summarizes the theoretical framework used to compute gravity-gradient signals relevant to CHRONOS. We consider three primary sources of terrestrial gravity perturbations: Rayleigh-wave seismic fields, atmospheric density fluctuations, and transient mass redistribution associated with earthquakes. Although their physical origins differ, all three processes can be described within a common framework based on density perturbations and their resulting gravitational fields.

The objective of this section is to establish the general relationship between environmental density fluctuations and the detector response. In particular, we derive the gravity-gradient tensor generated by an arbitrary density perturbation and provide the basis for its subsequent conversion into torsional motion of the detector.

The overall procedure can be summarized as

\begin{equation}
\delta\rho
\rightarrow
\delta\Phi
\rightarrow
\Gamma_{ij}
\rightarrow
\tau
\rightarrow
\theta ,
\end{equation}

where $\delta\rho$ is the density perturbation in the surrounding medium, $\delta\Phi$ is the resulting perturbation of the Newtonian gravitational potential, $\Gamma_{ij}$ is the gravity-gradient tensor, $\tau$ is the torque induced on the torsion bar, and $\theta$ is the measured angular displacement.

\subsection{Gravity gradient}

We first derive the gravity-gradient tensor produced by an arbitrary density perturbation $\delta\rho(\mathbf{x},t)$. The specific forms of $\delta\rho$ corresponding to seismic, atmospheric, and earthquake sources will be discussed in the following subsections.

The perturbation of the Newtonian gravitational potential at position $\mathbf{x}$ is

\begin{equation}
\delta\Phi(\mathbf{x},t)
=
-G
\int d^3x'
\frac{\delta\rho(\mathbf{x}',t)}
{|\mathbf{x}-\mathbf{x}'|},
\label{eq:potential}
\end{equation}

from which the perturbation of the gravitational acceleration is

\begin{equation}
\delta g_i(\mathbf{x},t)
=
-\partial_i \delta\Phi(\mathbf{x},t).
\label{eq:accel}
\end{equation}

The gravity-gradient tensor is then defined as

\begin{equation}
\Gamma_{ij}(\mathbf{x},t)
=
\partial_j \delta g_i
=
-\partial_i\partial_j \delta\Phi ,
\label{eq:gg}
\end{equation}

which describes the spatial variation of the gravitational acceleration field.

Using Eq.~(\ref{eq:potential}) and defining $\mathbf{R}=\mathbf{x}-\mathbf{x}'$, the gravity-gradient tensor can be written explicitly as

\begin{equation}
\Gamma_{ij}(\mathbf{x},t)
=
G
\int d^3x'\,
\delta\rho(\mathbf{x}',t)
\left[
\frac{\delta_{ij}}{R^3}
-
3\frac{R_iR_j}{R^5}
\right].
\label{eq:gg2}
\end{equation}

Equation~(\ref{eq:gg2}) provides the fundamental gravity-gradient field generated by an arbitrary environmental density perturbation. For CHRONOS, this tensor acts as the external driving field that produces differential gravitational forces on the end test masses and ultimately generates a measurable torsional response. In the following subsection, we derive the coupling between the gravity-gradient tensor and the torsional degree of freedom of the detector.

\subsection{Gravity gradient to detector signal}

This section describes the general conversion of an external gravity-gradient field into the measurable angular response of the CHRONOS torsion bar.

All gravity-gradient sources considered in this work, including Rayleigh-wave Newtonian noise, atmospheric Newtonian noise, and transient earthquake signals, couple to the detector through the same mechanical response. Once the gravity-gradient tensor $\Gamma_{ij}$ is known, the corresponding detector response can be obtained through the formalism developed below.

In the presence of an external gravity-gradient field, the torsion bar experiences a torque induced by $\Gamma_{ij}$, where $i,j\in\{x,y,z\}$ denote spatial directions.

Letting $\rho_{\rm TM}(\mathbf r)$ be the mass density distribution of the torsion bar, $\epsilon_{kab}$ the Levi--Civita symbol, and $\mathbf r$ the displacement from the center of mass, the torque is

\begin{equation}
\tau_k(t)
=
\int d^3r\,
\rho_{\rm TM}(\mathbf r)
\epsilon_{kab}
r_a
\delta g_b(\mathbf r,t).
\end{equation}

Expanding the gravitational acceleration about the center of mass and neglecting higher-order terms,

\begin{equation}
\delta g_b(\mathbf r,t)
\simeq
\delta g_b(\mathbf 0,t)
+
\Gamma_{bc}(t)r_c
+
\mathcal{O}(r^2),
\end{equation}

the first term produces no net torque about the center of mass. The dominant contribution is therefore

\begin{equation}
\tau_k(t)
=
\epsilon_{kab}
\Gamma_{bc}(t)
I_{ac},
\end{equation}

where

\begin{equation}
I_{ac}
=
\int d^3r\,
\rho_{\rm TM}(\mathbf r)
r_a r_c
\end{equation}

is the second moment tensor of the mass distribution.

For a cross-torsion bar aligned with the $x$ and $y$ axes, the off-diagonal components of the inertia tensor are negligible. The torque about the vertical axis then becomes

\begin{equation}
\tau_z(t)
=
(I_{xx}-I_{yy})
\Gamma_{xy}(t).
\label{eq:torque}
\end{equation}

Equation~(\ref{eq:torque}) shows that CHRONOS is primarily sensitive to the off-diagonal gravity-gradient component $\Gamma_{xy}$.

Combining Eqs.~(\ref{eq:torsion_susceptibility}) and Eqs.~(\ref{eq:torque}), the angular response to an arbitrary gravity-gradient source $X$ is

\begin{equation}
\theta_X(f)
=
\chi_\theta(f)
(I_{xx}-I_{yy})
\Gamma_{xy}^{X}(f).
\label{eq:thetaX}
\end{equation}

Equation~(\ref{eq:thetaX}) provides the general transfer function between an external gravity-gradient field and the measurable torsion-bar rotation. The following sections derive the source-dependent gravity-gradient fields $\Gamma_{xy}^{X}$ and apply Eq.~(\ref{eq:thetaX}) to calculate the corresponding detector response.

\subsection{Rayleigh geometric coupling}

\begin{figure}[H]
\centering
\includegraphics[width=0.45\linewidth]{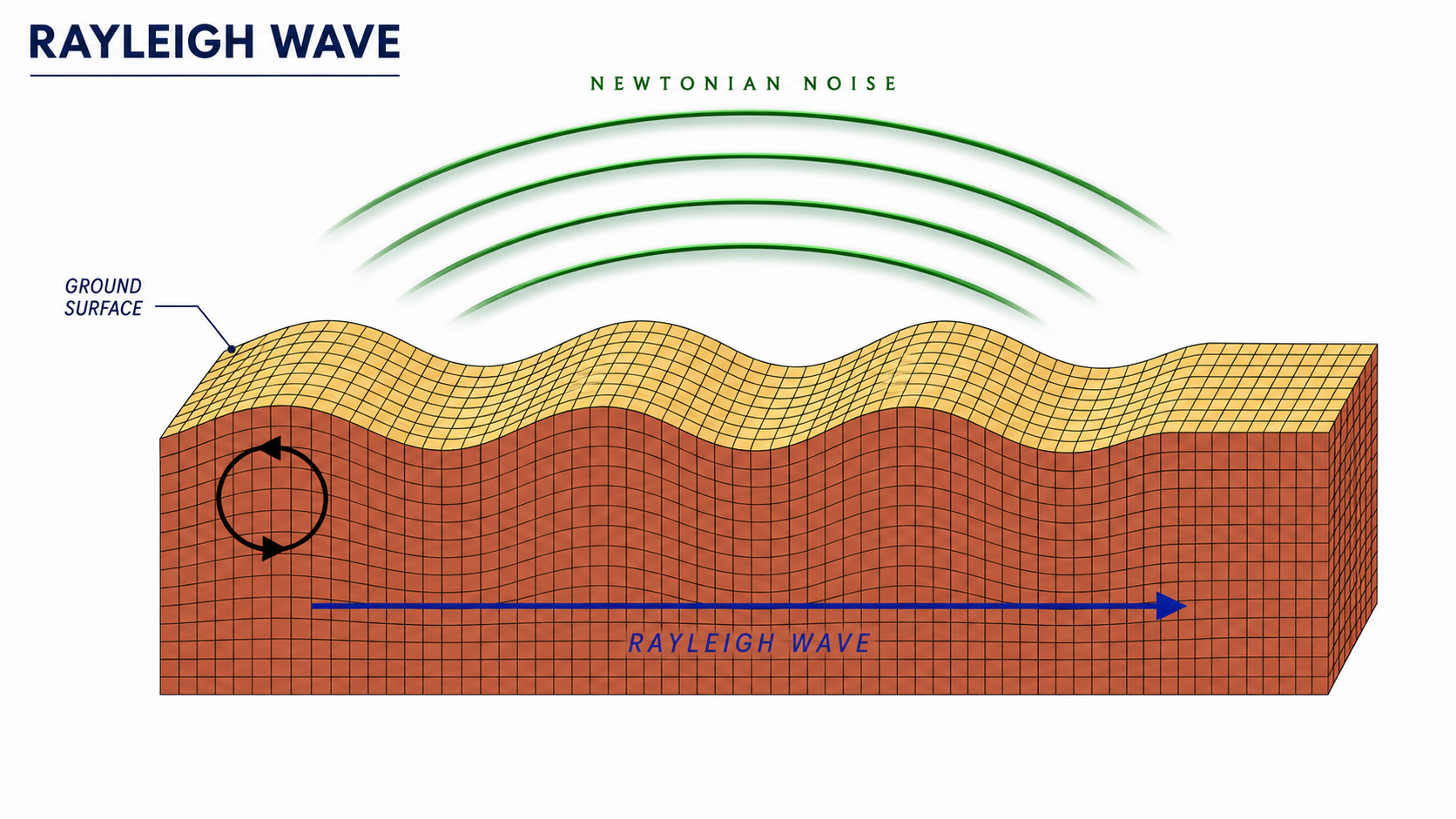}
\caption{Elliptical ground displacement induced by Rayleigh waves lead to ground mass density fluctuations and, thus, gravity perturbations. This couples gravitationally to the test mass mirrors, introducing noise to the readout.}
\label{fig:NN_concept}
\end{figure}

Rayleigh waves, which dominate seismic surface motion at low frequencies, produce gravity perturbations through ground density changes \cite{harms2013low} that are most prominent near the surface, as depicted in Fig.~\ref{fig:NN_concept}. These surface waves can be produced by multiple unpredictable sources, leading to a stochastic and stationary background, to a good approximation.

We model Rayleigh-wave Newtonian noise within the gravity-gradient framework developed in Sec.~\ref{sec:GG}. The density perturbation associated with a seismic displacement field $\boldsymbol{\xi}(\mathbf{r},t)$ is

\begin{equation}
\delta\rho_{\rm RW}
=
-\rho_0
\nabla\cdot
\left(
\eta\boldsymbol{\xi}
\right),
\end{equation}

where $\rho_0$ is the mean ground density and $\eta$ is a phenomenological quietness factor describing site-dependent suppression of seismic gravity-gradient noise.

Substituting Eq.~(\ref{eq:gg2}) into the gravity-gradient formalism yields the gravity-gradient tensor generated by Rayleigh waves. Rather than evaluating the full integral numerically, we adopt the analytical treatment of Ref.~\cite{harms2013low}, which gives the gravity-gradient field associated with a measured vertical ground displacement spectrum $\xi_z(f)$.

The dominant gravity-gradient component relevant to CHRONOS is

\begin{equation}
\Gamma_{xy}^{\rm RW}(f)
=
A
\left(
\frac{2 \pi f}{c_R}
\right)
\Gamma_x(f),
\end{equation}

where

\begin{equation}
\Gamma_x(f)
=
2\pi
G\rho_0
\gamma_R
\eta
\,
\xi_z(f)
\exp\!\left(
-\frac{2 \pi f h}{c_R}
\right),
\end{equation}

and

\begin{equation}
A
=
\sqrt{\langle\cos^4\theta\rangle}
=
\sqrt{\frac{3}{8}}
\end{equation}

is the RMS angular averaging factor. Here $c_R$ is the Rayleigh-wave velocity, $h$ is the detector height above the ground, and $\gamma_R\simeq0.83$ accounts for partial cancellation between compressional and shear components of the seismic field.

Using the coupling derived in Sec.~\ref{sec:GG}, the gravity-gradient field generates a torque

\begin{equation}
\tau_z(f)
=
(I_{xx}-I_{yy})
\Gamma_{xy}^{\rm RW}(f).
\label{eq:Rayleigh_torque}
\end{equation}

The resulting angular response is obtained using the torsional susceptibility introduced in Sec.~\ref{sec:CHRONOS},

\begin{equation}
\theta_{\rm RW}(f)
=
\chi_\theta(f)
\tau_z(f).
\label{eq:Rayleigh_angle}
\end{equation}

Combining the above expressions gives the complete transfer function

\begin{equation}
\theta_{\rm RW}(f)
=
\chi_\theta(f)
(I_{xx}-I_{yy})
\Gamma_{xy}^{\rm RW}(f),
\end{equation}

which directly relates Rayleigh-wave-induced gravity gradients to the measurable angular motion of the CHRONOS torsion bar.

Since Rayleigh-wave gravity gradients constitute a stationary stochastic background, they are characterized by power spectral densities. Using Eq.~(\ref{eq:thetaX}), the corresponding angular-noise spectrum is

\begin{equation}
S_{\theta}^{\rm RW}(f)
=
\left|
\chi_\theta(f)
(I_{xx}-I_{yy})
\right|^2
S_{\Gamma_{xy}}^{\rm RW}(f),
\label{eq:S_theta_RW}
\end{equation}

where $S_{\Gamma_{xy}}^{\rm RW}(f)$ is the power spectral density of the Rayleigh-wave gravity-gradient field.

To compare the resulting gravity-gradient noise directly with the detector strain sensitivity, the angular spectrum is converted into an equivalent gravitational-wave strain spectrum using the gravitational-wave coupling coefficient $\eta_g=0.936$. This coefficient relates the incident gravitational-wave strain to the torsional response of CHRONOS and was derived in Inoue et al.~\cite{inoue2026improvingcalibrationaccuracytorque} from the gravitational coupling of the detector geometry.

\begin{equation}
S_h^{\rm RW}(f)
=
\frac{
S_{\theta}^{\rm RW}(f)
}
{
|\eta_g|^2
}.
\label{eq:S_h_RW}
\end{equation}

Equation~(\ref{eq:S_h_RW}) defines the strain-equivalent Rayleigh-wave Newtonian-noise spectrum used throughout the remainder of this work.

\subsection{Atmospheric coupling}

Atmospheric or infrasound NN arises from pressure fluctuations that induce density variations in the air surrounding the detector, leading to time-varying gravitational fields that couple directly to the test masses. For low-frequency ground-based detectors, it has been predicted that atmospheric coupling is orders of magnitude smaller than the targeted sensitivity of around $10^{-18}\,\mathrm{Hz^{-1/2}}$ near $0.1~\mathrm{Hz}$ \cite{fiorucci2018impact}.

Within the framework introduced in Sec.~\ref{sec:GG}, atmospheric pressure fluctuations act as a source of density perturbations and therefore generate gravity-gradient fields that couple to the torsional degree of freedom of CHRONOS.

The density perturbation of an infrasound wave is given by

\begin{equation}
\delta \rho_{\rm ATM}
=
\frac{\rho_0}{\gamma}
\frac{\delta p}{p_0},
\label{eq:Atmospheric}
\end{equation}

where $\rho_0$ is the mean air density, $\gamma$ is the adiabatic coefficient of air, and $\delta p/p_0$ is the relative pressure fluctuation.

Equation~(\ref{eq:Atmospheric}) follows from the adiabatic relation between pressure and density perturbations in a propagating infrasound wave.

To estimate the detector sensitivity, the atmospheric pressure spectrum is approximated by a phenomenological broken power law (Eq.~(\ref{eq:pressurefluctuations})) with a break frequency of $f_0=0.1~\mathrm{Hz}$, motivated by the transition to the microbarom-dominated regime \cite{10.1093/gji/ggaa015}. Broken power-law descriptions are commonly used for atmospheric fluctuation spectra, although the exact spectral indices, $\alpha_1$ and $\alpha_2$, depend on the underlying physical processes and observational conditions \cite{waite2020untangling}. For this work, a simplified model is used where $f_0=0.1~\mathrm{Hz}$, $\alpha_1=1$, and $\alpha_2=2$. The chosen exponents are not intended to represent a specific atmospheric model, but rather to capture the existence of different spectral scaling regimes observed in atmospheric and infrasound measurements.

\begin{equation}
\frac{\delta p(f)}{p_0}
=
\begin{cases}
\left(\dfrac{f}{f_0}\right)^{-\alpha_1},
&
f<f_0,
\\[0.2cm]
\left(\dfrac{f}{f_0}\right)^{-\alpha_2},
&
f\ge f_0,
\end{cases}
\label{eq:pressurefluctuations}
\end{equation}

Substituting Eq.~(\ref{eq:Atmospheric}) into the gravity-gradient formalism and adopting the analytical treatment of Ref.~\cite{harms2013low}, the atmospheric gravity perturbation can be expressed as

\begin{equation}
x(f)
=
-4\pi i
\sin \phi
\cos \theta
G
\delta \rho_{\rm ATM}
c_{IS}
\exp
\left(
\frac{\sin \phi \, 2 \pi f z_0}
     {c_{IS}}
\right)
(2 \pi f)^{-3},
\label{eq:Atmospheric_disp}
\end{equation}

where $\phi$ is the angle of incidence of the infrasound wave with respect to the normal of Earth's surface, $\theta$ is the angle between the detector axis and the horizontal propagation direction of the wave, $c_{IS}$ is the infrasound wave speed, and $z_0\le0$ is the depth of the detector relative to the surface.

From this expression, it can be observed that the exponential suppression with depth strongly depends on the incident angle. Nearly horizontal infrasound waves have a large projection onto the detector axis and can therefore be efficiently suppressed by underground installation, whereas vertically propagating waves exhibit a weaker dependence on depth.

The corresponding gravitational acceleration is

\begin{equation}
g_x(f)
=
-(2 \pi f)^2 x(f).
\end{equation}

Approximating the infrasound field as a plane wave with characteristic wavenumber

\begin{equation}
k_{IS}
=
\frac{2 \pi f}{c_{IS}},
\end{equation}

the associated off-diagonal gravity-gradient component can be estimated from the spatial variation of the gravitational acceleration as

\begin{equation}
\Gamma_{xy}^{\rm ATM}(f)
=
A
k_{IS}
g_x(f),
\label{eq:Atmospheric_Gamma}
\end{equation}

where

\begin{equation}
A
=
\sqrt{\langle\cos^4\theta\rangle}
=
\sqrt{\frac{3}{8}}
\end{equation}

is the RMS angular averaging factor for an isotropic distribution of infrasound propagation directions.

Using Eq.~(\ref{eq:thetaX}), the angular response induced by atmospheric gravity gradients is

\begin{equation}
\theta_{\rm ATM}(f)
=
\chi_\theta(f)
\left(
I_{xx}
-
I_{yy}
\right)
\Gamma_{xy}^{\rm ATM}(f).
\label{eq:Atmos_angle}
\end{equation}

Since atmospheric Newtonian noise is treated as a stationary stochastic process, it is characterized by a power spectral density. The corresponding angular-noise spectrum is

\begin{equation}
S_{\theta}^{\rm ATM}(f)
=
\left|
\chi_\theta(f)
\left(
I_{xx}
-
I_{yy}
\right)
\right|^2
S_{\Gamma_{xy}}^{\rm ATM}(f),
\label{eq:S_theta_ATM}
\end{equation}

where $S_{\Gamma_{xy}}^{\rm ATM}(f)$ is the atmospheric gravity-gradient power spectral density.

The corresponding strain-equivalent Newtonian-noise spectrum is obtained through the gravitational-wave response function $\eta_g(f)$,

\begin{equation}
S_h^{\rm ATM}(f)
=
\frac{
S_{\theta}^{\rm ATM}(f)
}
{
|\eta_g|^2
}.
\label{eq:S_h_ATM}
\end{equation}

Equation~(\ref{eq:S_h_ATM}) defines the strain-equivalent atmospheric Newtonian-noise spectrum used in the following sensitivity calculations.

\subsection{Transient seismic coupling}

\begin{figure}[H]
\centering
\includegraphics[width=0.45\linewidth]{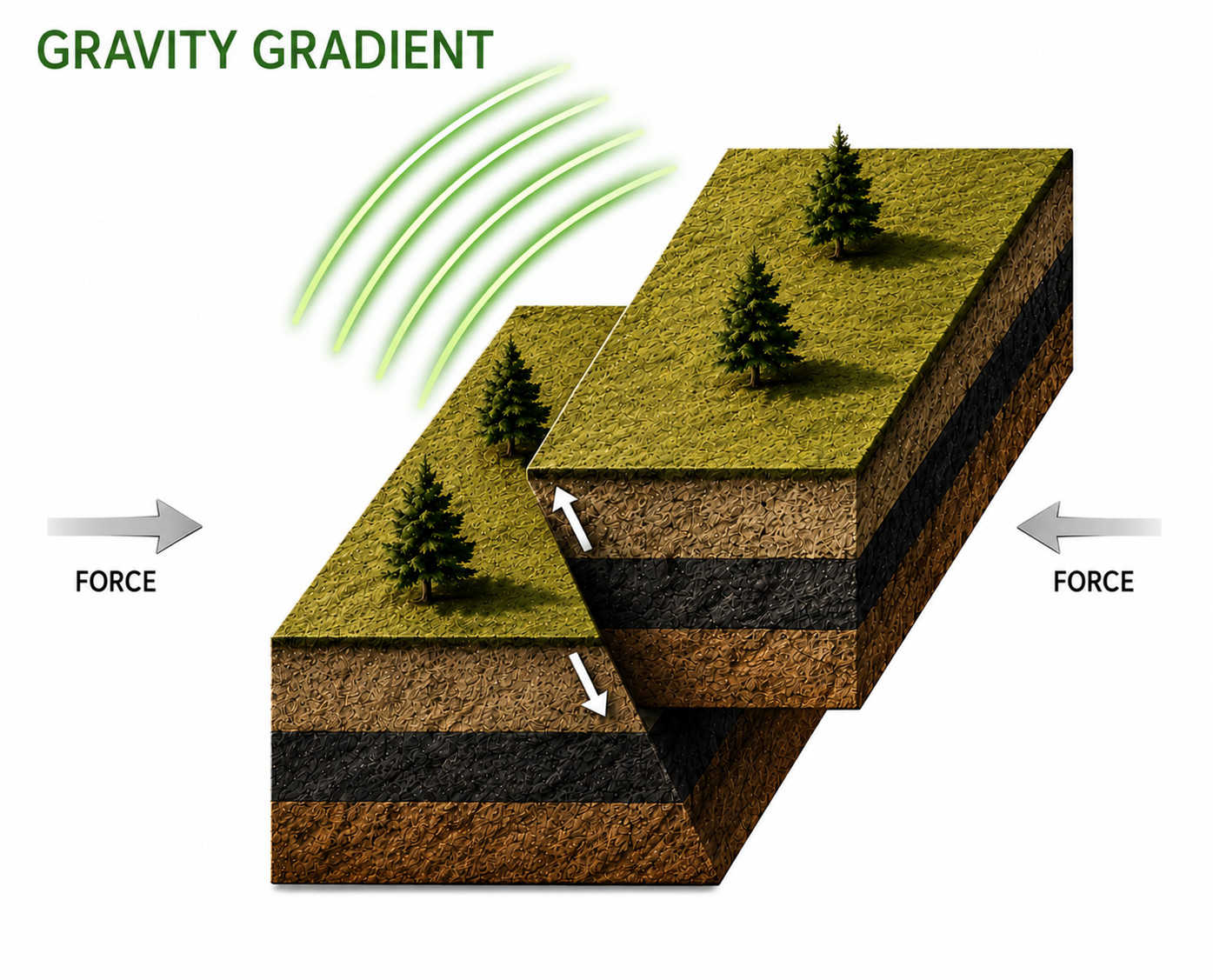}
\caption{Drastic changes in ground distribution due to earthquakes produce a multitude of body waves and surface waves. Aside from these, the density perturbation itself produces a change in the gravitational field that then couples to the detector.}
\label{fig:EQ_concept}
\end{figure}

Rapid mass redistribution during an earthquake generates time-dependent gravitational perturbations that propagate at the speed of light, preceding the arrival of seismic waves, as visualized in Fig.~\ref{fig:EQ_concept}.

Within the framework introduced in Sec.~\ref{sec:GG}, earthquakes act as transient sources of density perturbations. Unlike Rayleigh-wave and atmospheric Newtonian noise, these perturbations are coherent transient events associated with rapid redistribution of underground mass.

Density fluctuations associated with seismic events can be obtained from the seismic potential $\phi_s$ by \cite{harms2015transient}

\begin{equation}
\delta\rho_{EQ}
=
-\rho_0\nabla^2\phi_s.
\end{equation}

The resulting perturbation of the gravitational field can be expressed in terms of the seismic moment time function. Following Ref.~\cite{harms2015transient}, the orientation-averaged prompt gravity signal is given by

\begin{equation}
\langle h_{+}(\mathbf{r}_0,t) \rangle
=
\langle h_{\times}(\mathbf{r}_0,t) \rangle
=
\frac{6\sqrt{14/5}\,G}{r_0^5}
I_4[M_0](t),
\label{eq:transient}
\end{equation}

where $r_0$ is the distance to the earthquake centroid and $I_4[M_0](t)$ is the fourth time integral of the seismic moment time function,

\begin{equation}
I_4[M_0](t)
=
\int_0^t dt'
\int_0^{t'} dt''
\int_0^{t''} dt'''
\int_0^{t'''} dt''''
M_0(t'''')
\equiv
I_4[M_0](t).
\label{eq:M0time}
\end{equation}

Although Eq.~(\ref{eq:transient}) is written in a strain-like form, it should be interpreted as a strain-equivalent representation of the prompt Newtonian gravity perturbation generated by the earthquake rather than as a propagating gravitational wave. In the following, we denote this quantity by $h_{EQ}(t)$.


\section{Results}
\label{sec:results}

\subsection{Rayleigh-wave and atmospheric gravity-gradient noise}

We present the stationary gravity-gradient background arising from ambient seismic and atmospheric fluctuations, obtained from measured seismic data at the proposed site using Eq.~(\ref{eq:Rayleigh_angle}), together with a simplified empirical model for atmospheric fluctuations using Eq.~(\ref{eq:Atmos_angle}).

\begin{figure}[H]
\centering
\includegraphics[width=0.7\linewidth]{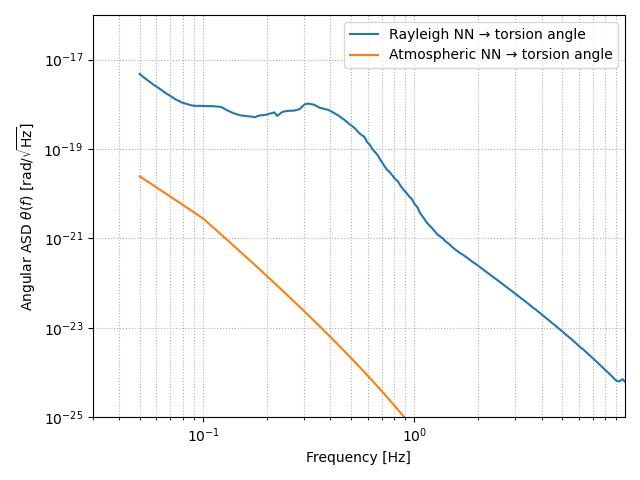}
\caption{Angular amplitude spectral density of gravity-gradient backgrounds from Rayleigh-wave (green) and atmospheric (orange) sources. The Rayleigh contribution dominates throughout the frequency band considered and exhibits a broad excess around 0.3~Hz associated with the secondary microseism band.}
\label{fig:NN}
\end{figure}

Figure~\ref{fig:NN} shows the resulting angular amplitude spectral densities of the two contributions. For the model parameters adopted in this study, the atmospheric contribution remains several orders of magnitude below the Rayleigh-wave contribution throughout the entire frequency range considered. Consequently, the predicted gravity-gradient background is dominated by seismic surface-wave coupling.

A notable feature of the Rayleigh spectrum is the broad excess centered around 0.3~Hz. This frequency range overlaps with the secondary microseism band, which is commonly observed between 0.1 and 0.4~Hz and is generated by the nonlinear interaction of ocean waves propagating in opposite directions. The presence of this feature suggests that regional ocean-wave activity may play an important role in determining the local gravity-gradient environment.

The dominance of the Rayleigh contribution implies that accurate characterization of the seismic environment is essential for predicting the gravity-gradient background at a candidate CHRONOS site. In contrast, uncertainties in the atmospheric model are expected to have a negligible impact on the overall background level for the detector configurations considered here.

The dependence of the gravity-gradient background on detector depth and seismic quietness is investigated in the following section.


\subsection{Dependence on detector depth and seismic quietness}
\begin{figure}[H]
\centering
\includegraphics[width=0.8\linewidth]{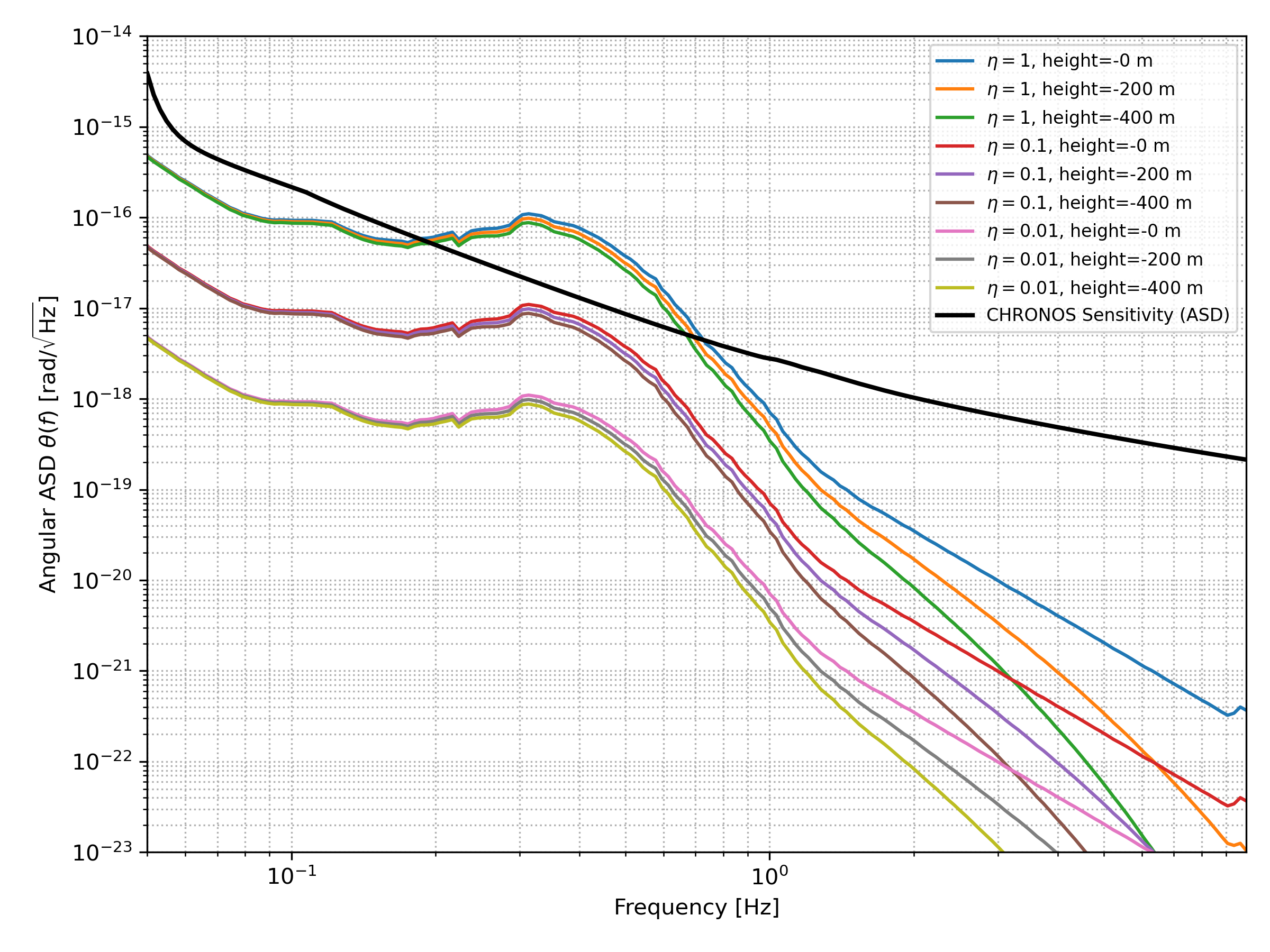}
\caption{Calculated gravity-gradient background spectra for different detector depths ($h$) and seismic quietness factors ($\eta$), compared with the projected CHRONOS sensitivity (black)~\cite{inoue2026opticaldesignsensitivityoptimization}. Variations in $\eta$ produce the largest changes in amplitude, while depth introduces a weaker frequency-dependent suppression.}
\label{fig:NN_multi}
\end{figure}

Figure~\ref{fig:NN_multi} shows the gravity-gradient background spectra for various assumed detector depths $h$ and seismic quietness factors $\eta$, calculated using Eq.~\eqref{eq:Rayleigh_angle}, and compared with the projected CHRONOS sensitivity.

The total strain sensitivity used for comparison is adopted from the optimized CHRONOS configuration presented in Inoue et al.~\cite{inoue2026opticaldesignsensitivityoptimization}. The sensitivity curve includes all major instrumental noise contributions, including quantum noise, coating thermal noise, torsion-bar thermal noise, and residual seismic noise. The gravity-gradient spectra calculated in this work are not included in the baseline sensitivity curve and are shown separately to evaluate their impact on detector performance.

These parameters were varied because they represent the dominant site-dependent effects governing Rayleigh-wave gravity-gradient backgrounds. The detector depth controls the attenuation of the surface-wave field, whereas the quietness factor parameterizes the overall level of ambient seismic activity.

It can be observed that the quietness factor has a more pronounced effect on the overall amplitude of the gravity-gradient background, while the depth induces smaller variations that become increasingly important toward higher frequencies.

For $\eta=1$, the predicted gravity-gradient background exceeds the nominal CHRONOS sensitivity between approximately $0.2~\mathrm{Hz}$ and $0.7~\mathrm{Hz}$ for all detector depths considered. In contrast, for smaller values of $\eta$, the background falls below the detector sensitivity over most of the frequency range.

This result demonstrates that the local seismic environment is the primary factor determining the significance of Rayleigh-wave gravity-gradient backgrounds. Although underground installation provides additional suppression, improvements in site quietness produce substantially larger reductions in the predicted background level.

Consequently, site selection and environmental characterization are expected to play a central role in future gravity-gradient measurements and low-frequency gravitational-wave detector development.

\newpage
\subsection{Detectability of prompt earthquake gravity signals}

\begin{figure}[H]
\centering
\includegraphics[width=0.75\linewidth]{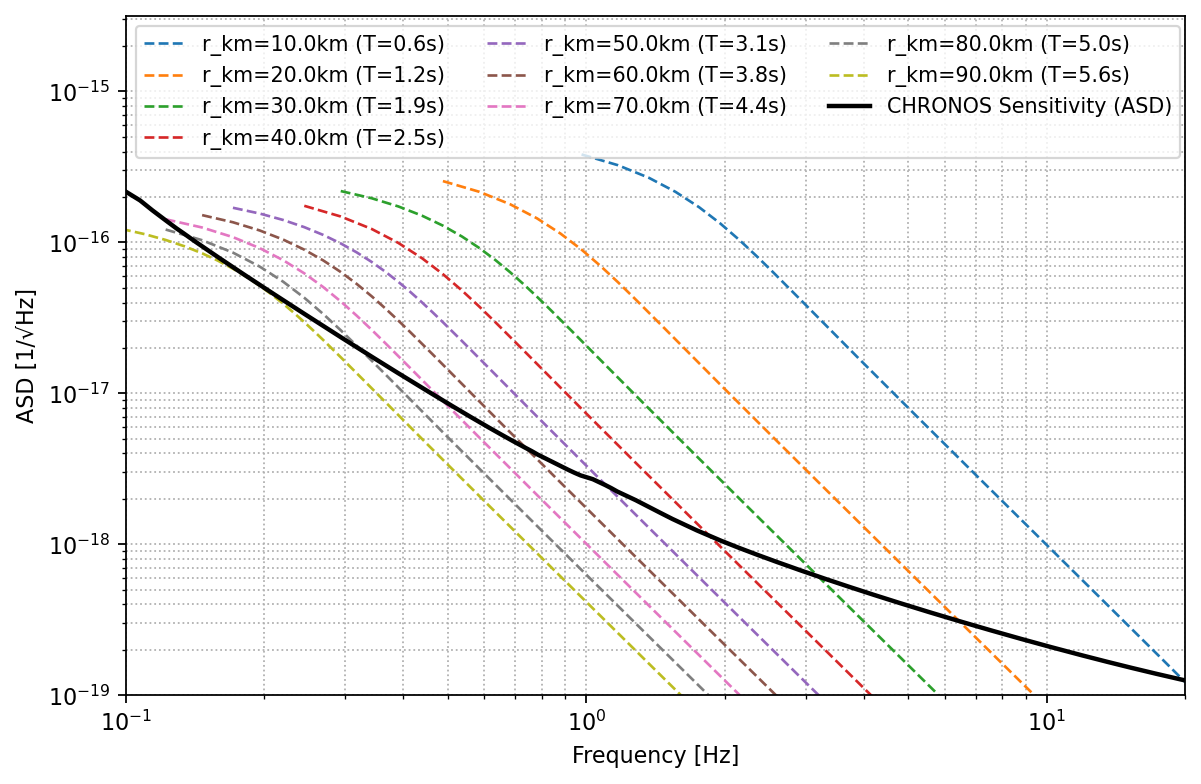}
\caption{Projected strain sensitivity of CHRONOS overlaid with strain-equivalent prompt gravity signals $h_{EQ}(f)$ from an $M_w=5.2$ earthquake at various source distances. T is half the duration of time between the arrival of the prompt gravity signals and the seismic P-waves. As source distance decreases, the signal increases in amplitude and shifts to higher frequencies.}
\label{fig:EQ_a}
\end{figure}

Figure~\ref{fig:EQ_a} compares the projected CHRONOS sensitivity with the strain-equivalent prompt gravity signals generated by a
representative $M_w=5.2$ earthquake at different source distances.

The signals correspond to a moment magnitude $M_w = 5.2$ earthquake, chosen as a representative lower threshold for engineering significance \cite{thenhaus1994estimates}. The spectra are obtained from Eq.~(\ref{eq:transient}) and are presented in terms of the strain-equivalent Fourier-domain amplitude $h_{EQ}(f)$.

As shown in Fig.~\ref{fig:EQ_a}, for source distances less than $90$\,km, the predicted signal lies within the CHRONOS sensitivity band in the sub-Hz regime, indicating the possibility of detection. The closer the source, the larger the signal amplitude and the broader the frequency range over which the signal exceeds the detector sensitivity. However, there is a trade-off in that the time between rupture onset and P-wave arrival becomes shorter at closer distances, corresponding to a smaller measurement window of the prompt gravitational signal, as indicated by increasing cut-off frequency.

To quantify detectability, we compute the signal-to-noise ratio (SNR) of the prompt earthquake gravity signal. The quantity $h_{EQ}(f)$ denotes the strain-equivalent Fourier-domain amplitude of the prompt gravity signal.

\begin{equation}
\mathrm{SNR}
=
\sqrt{
4
\int_{f_{\rm min}}^{f_{\rm max}}
\frac{
|h_{EQ}(f)|^2
}
{
S_n(f)
}
\,df
},
\label{eq:SNR}
\end{equation}

where $S_n(f)$ is the one-sided detector noise power spectral density.

\begin{figure}[H]
\centering
\includegraphics[width=0.45\linewidth]{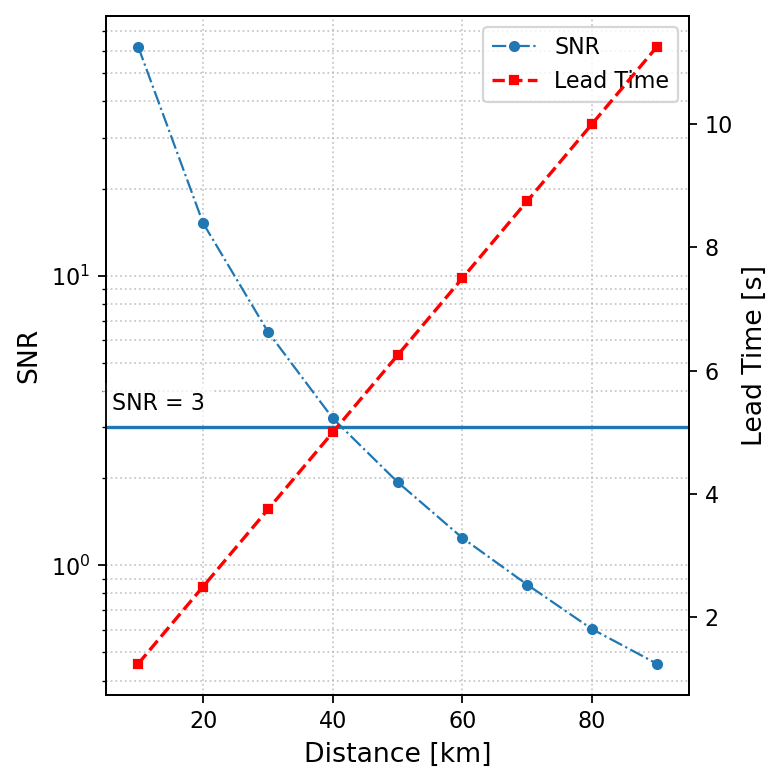}
\caption{SNRs and lead time of an $M_w=5.2$ earthquake at various source distances. Increasing distance extends the duration between P-wave and prompt gravity signal arrival, increasing lead time while simultaneously reducing signal strength.}
\label{fig:EQ_b}
\end{figure}

Figure~\ref{fig:EQ_b} shows the resulting SNR together with the corresponding lead time, defined as the time difference between the arrival of the prompt gravitational signal and the seismic P-wave. An average P-wave velocity of $8.0$\,km\,s$^{-1}$ is assumed. The results show that the SNR decreases with increasing distance, while the lead time increases due to the finite propagation speed of seismic waves. At a representative distance of $40$\,km, we obtain $\mathrm{SNR} \approx 3.62$ and a lead time of approximately $5.0$\,s. Considering a range of P-wave velocities between $5.8$ and $13.7$\,km\,s$^{-1}$ \cite{bormann2012seismic}, the corresponding lead time varies between $2.9$ and $6.9$\,s.

These results indicate that sufficiently nearby earthquakes could produce observable prompt gravity perturbations prior to seismic-wave arrival.

\begin{figure}[H]
\centering
\includegraphics[width=0.6\linewidth]{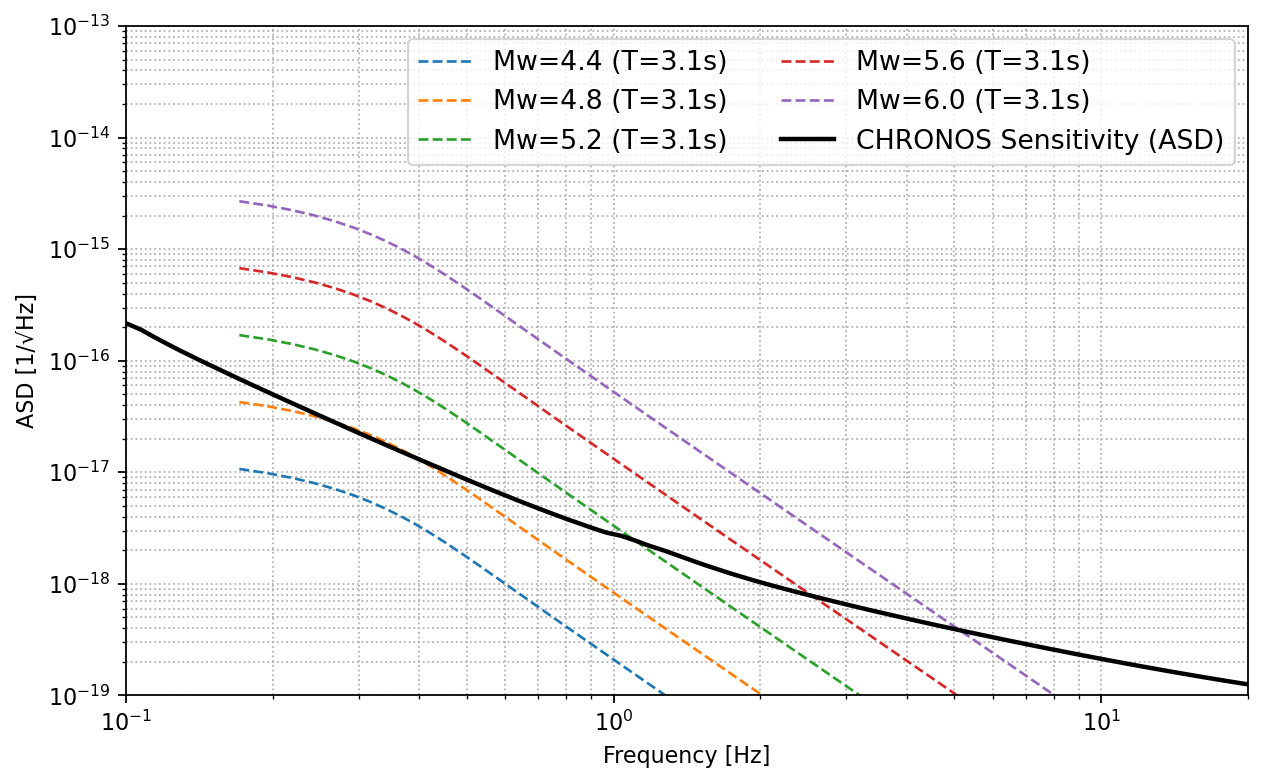}
\caption{Projected strain sensitivity of CHRONOS overlaid with strain-equivalent prompt gravity signals $h_{EQ}(f)$ from earthquakes of various moment magnitudes at a distance of $50~\mathrm{km}$.}
\label{fig:EQ_a2}
\end{figure}

Figure~\ref{fig:EQ_a2} contains the projected CHRONOS sensitivity together with simulated prompt gravitational signals from earthquakes at a fixed source distance of $50~\mathrm{km}$ and varying moment magnitude from $4.2$ to $6.0$.

As expected, increasing moment magnitude primarily changes the signal amplitude while leaving the overall spectral shape largely unchanged. The lead time remains unchanged because it depends only on the source distance.

\begin{figure}[H]
\centering
\includegraphics[width=0.45\linewidth]{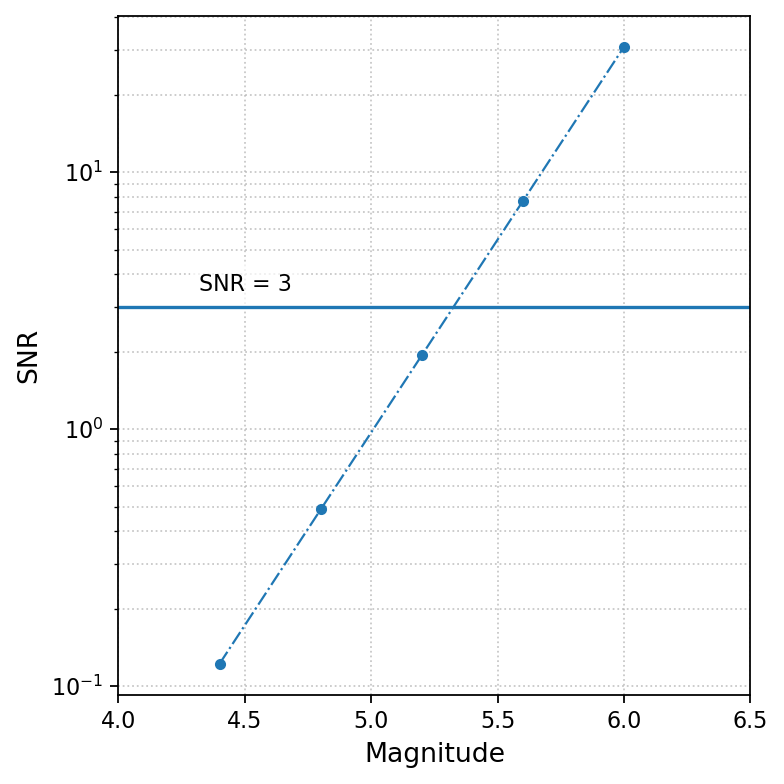}
\caption{SNRs of various earthquake magnitudes at a fixed source distance of $50$\,km. The SNR increases with increasing moment magnitude, and the SNR = 3 threshold is reached at approximately $M_w=5.3$.}
\label{fig:EQ_b2}
\end{figure}

Figure~\ref{fig:EQ_b2} shows the resulting SNR for various earthquake moment magnitudes at a representative distance of $50$\,km. As expected from the scaling of seismic moment with moment magnitude, the observed SNR increases rapidly with increasing magnitude over the range considered. The horizontal reference line at SNR = 3 is used as a detection threshold, indicating the minimum level at which a signal becomes distinguishable from the detector noise. From the trend, this threshold is crossed at approximately moment magnitude $5.3$ for a $50$\,km source distance.

\begin{figure}[H]
\centering
\includegraphics[width=0.6\linewidth]{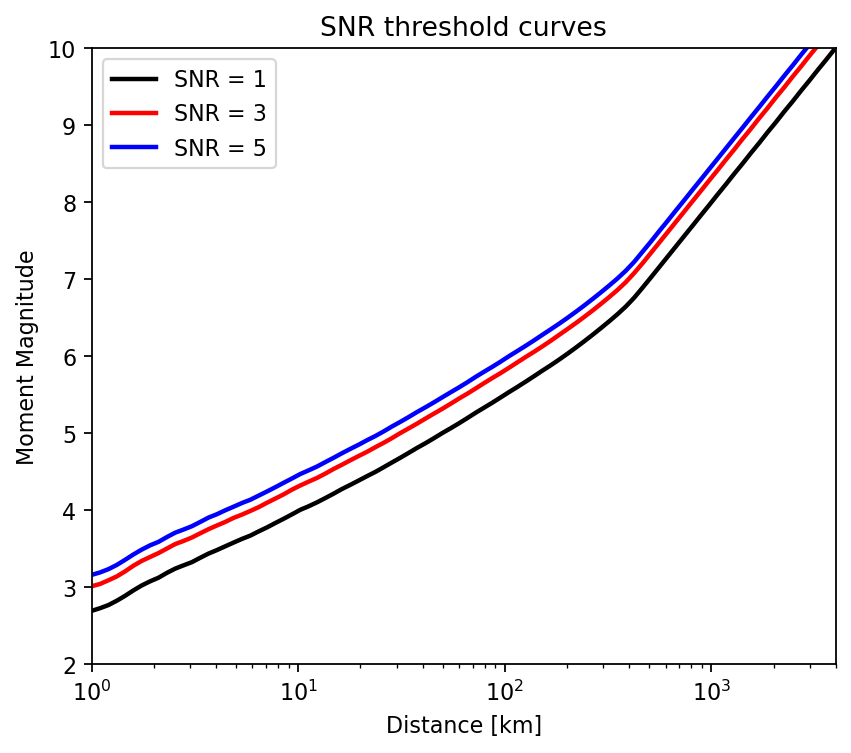}
\caption{Minimum detectable moment magnitude as a function of source distance for SNR thresholds of 1, 3, and 5. Regions above and to the left of each curve correspond to detectable events.}
\label{fig:EQ_c}
\end{figure}

Figure~\ref{fig:EQ_c} provides a convenient summary of the accessible distance--magnitude parameter space for the prototype CHRONOS detector. Figure~\ref{fig:EQ_c} illustrates the minimum detectable moment magnitude as a function of source--site distance for three SNR thresholds (SNR = 1, 3, and 5) obtained by inversion of Eq.~(\ref{eq:SNR}). Each curve represents the detection limit under the corresponding SNR criterion, with higher SNR thresholds requiring larger moment magnitudes at a given distance.

It is observed that the detection boundary shifts toward higher magnitudes with increasing distance, reflecting the geometric attenuation of the prompt gravity signal. Equivalently, for a fixed magnitude, the achievable SNR decreases with increasing distance. Regions above and to the left of each curve correspond to detectable events, with more favorable detection conditions attained for larger magnitudes and shorter source distances.


\section{Discussion}
\label{sec:discussion}

\subsection{Site dependence of gravity-gradient backgrounds} The calculated gravity-gradient background is dominated by Rayleigh-wave induced density perturbations, while the atmospheric contribution remains several orders of magnitude smaller for the model parameters considered. This behavior is consistent with previous studies of terrestrial gravity-gradient noise in the sub-Hz regime, where seismic surface waves are expected to provide the dominant environmental contribution. A notable feature of the calculated Rayleigh spectrum is the broad excess around 0.3~Hz shown in Fig.~\ref{fig:NN}. This frequency range overlaps with the secondary microseism band, which is generated by the nonlinear interaction of ocean waves propagating in opposite directions \cite{tanimoto2023seismic,longuet1950theory}. The result suggests that local geography and proximity to coastlines may significantly influence the gravity-gradient environment experienced by a detector. The dependence on detector depth and seismic quietness further demonstrates that site conditions play a central role in determining the achievable gravity-gradient background. While increasing depth reduces the coupling to surface Rayleigh waves, variations in the seismic quietness factor produce substantially larger changes in the overall amplitude. These results indicate that site characterization may be at least as important as detector design when targeting gravity-gradient measurements in the sub-Hz band. The Rayleigh and atmospheric contributions considered here should be interpreted as stochastic gravity-gradient backgrounds rather than deterministic signals. Unlike earthquake-induced perturbations discussed below, these fluctuations arise from stationary environmental processes and are therefore characterized statistically through their power spectral densities. 

\subsection{Prospects for gravity-based earthquake detection}

In contrast to the stationary gravity-gradient backgrounds discussed above, earthquake-induced perturbations are transient events associated with specific geophysical sources rather than stochastic environmental processes. For this reason, they are treated separately from the stochastic background contributions considered in this work.

The simulations presented in Sec.~\ref{sec:results} indicate that CHRONOS may be capable of detecting prompt gravity perturbations prior to the arrival of seismic P waves. For a representative earthquake with moment magnitude $M_w = 5.2$, detectable signals are predicted for source distances below approximately 90~km.

Unlike conventional seismic early-warning systems, which rely on elastic waves propagating through the Earth, gravity-based detection probes changes in the mass distribution itself. Consequently, the gravitational perturbation propagates at the speed of light and is not limited by seismic-wave travel times. Although the obtainable warning times are modest for nearby earthquakes, the method provides information that is fundamentally independent of seismic-wave observations and could therefore complement existing seismic monitoring networks.

More broadly, the ability to observe prompt gravity perturbations would provide a direct probe of earthquake-induced mass redistribution. Such measurements could offer a new observational channel for studying the earliest stages of earthquake rupture and the associated evolution of the seismic moment.

\subsection{Implications from historical earthquakes}

The practical significance of the predicted earthquake gravity signals can be assessed by comparing the estimated detection region with the historical seismicity around Taiwan. To this end, earthquakes with moment magnitude $M_w \ge 5$ and source--site distances smaller than 100~km occurring between January 1995 and May 2026 were extracted from the Central Weather Administration (CWA) earthquake catalog \cite{CWA_EarthquakeCatalog}.

\begin{figure}[H]
     \centering
     \begin{subfigure}{0.38\textwidth}
         \centering
         \includegraphics[width=\linewidth]{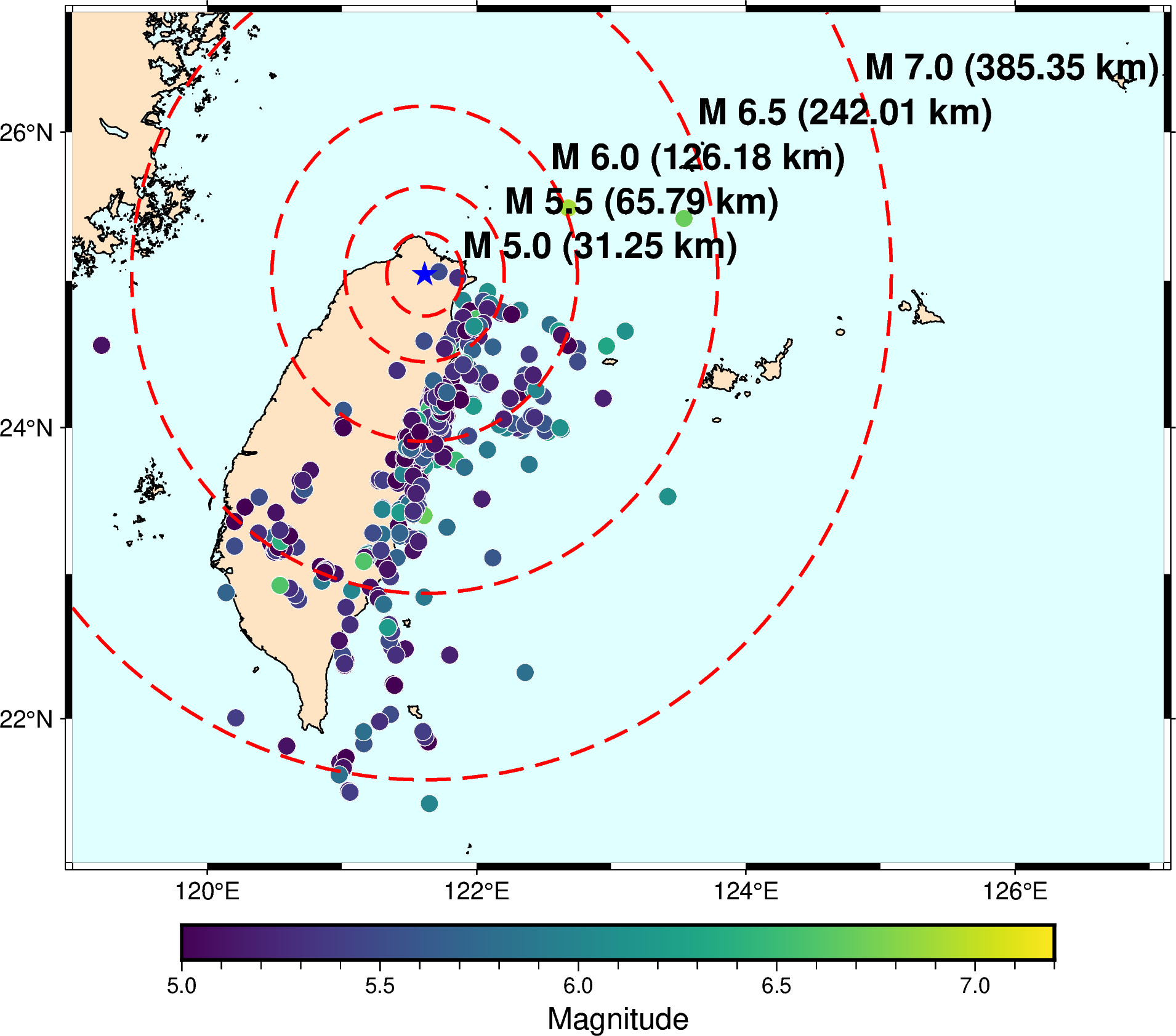}
         \caption{}
         \label{fig:EQ_map}
     \end{subfigure}
     \begin{subfigure}{0.56\textwidth}
         \centering
         \includegraphics[width=\linewidth]{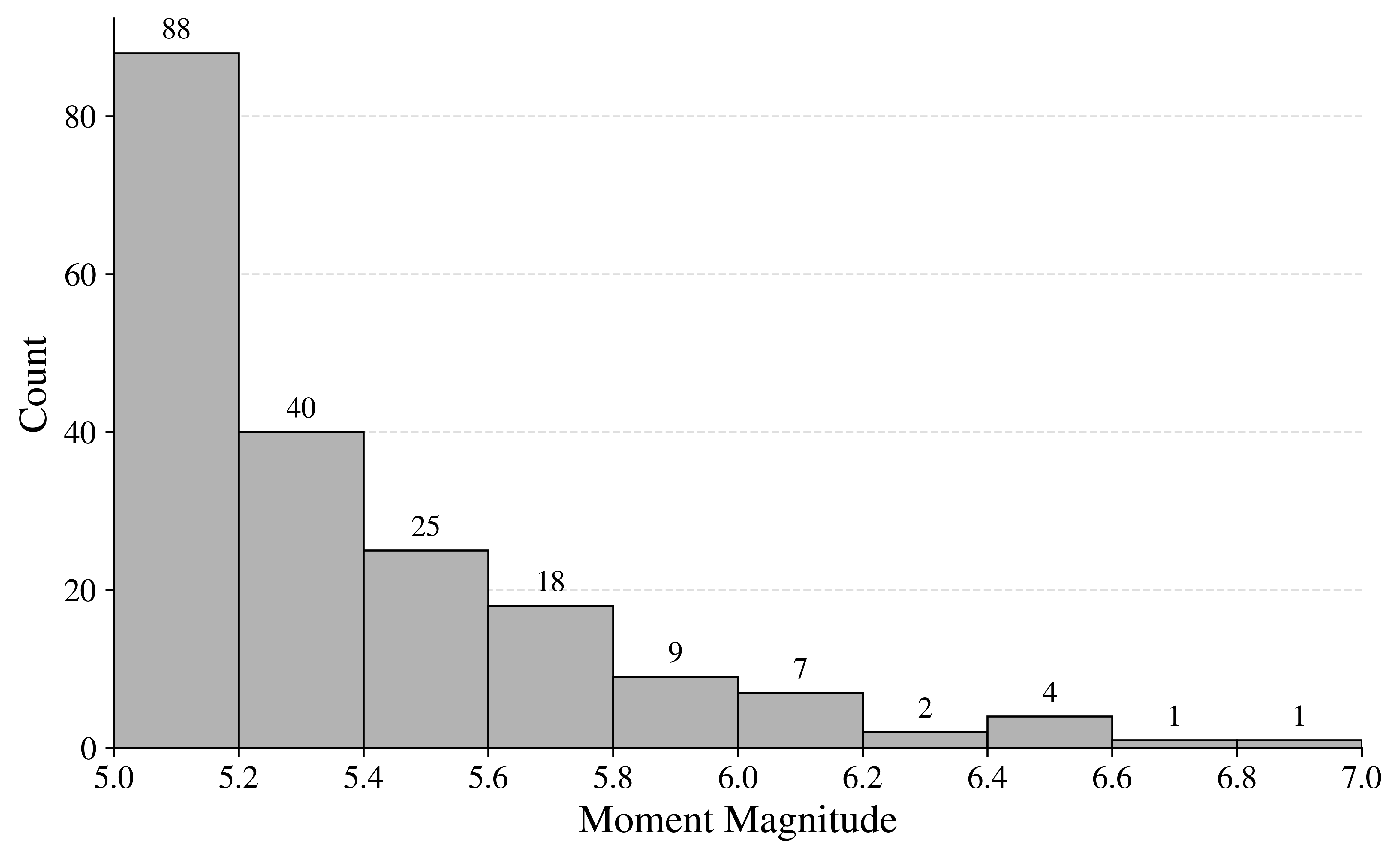}
         \caption{}
         \label{fig:EQ_hist}
     \end{subfigure}
     \caption{(a) Historical earthquakes with moment magnitudes $M_w \ge 5$ from the past 10 years with rings corresponding to SNR = 3, and (b) the corresponding magnitude distribution from January 1995 to May 2026 \cite{CWA_EarthquakeCatalog}.}
     \label{fig:EQ_maphist}
\end{figure}

The corresponding events are shown in Fig.~\ref{fig:EQ_map}, while the distribution of earthquake magnitudes is summarized in Fig.~\ref{fig:EQ_hist}.

Within this selection, the historical record corresponds to an average occurrence rate of approximately 6.3 events per year. Most events fall within the magnitude range $5 \le M_w \le 6$, indicating that moderate earthquakes dominate the local seismic activity relevant to CHRONOS.

The occurrence rate inferred from the historical catalog suggests that candidate events are not exceptionally rare. Consequently, if the projected detector sensitivity is achieved, repeated observations of earthquake-induced gravity perturbations may be possible over the operational lifetime of the detector.

\begin{figure}[H]
\centering
\includegraphics[width=0.6\linewidth]{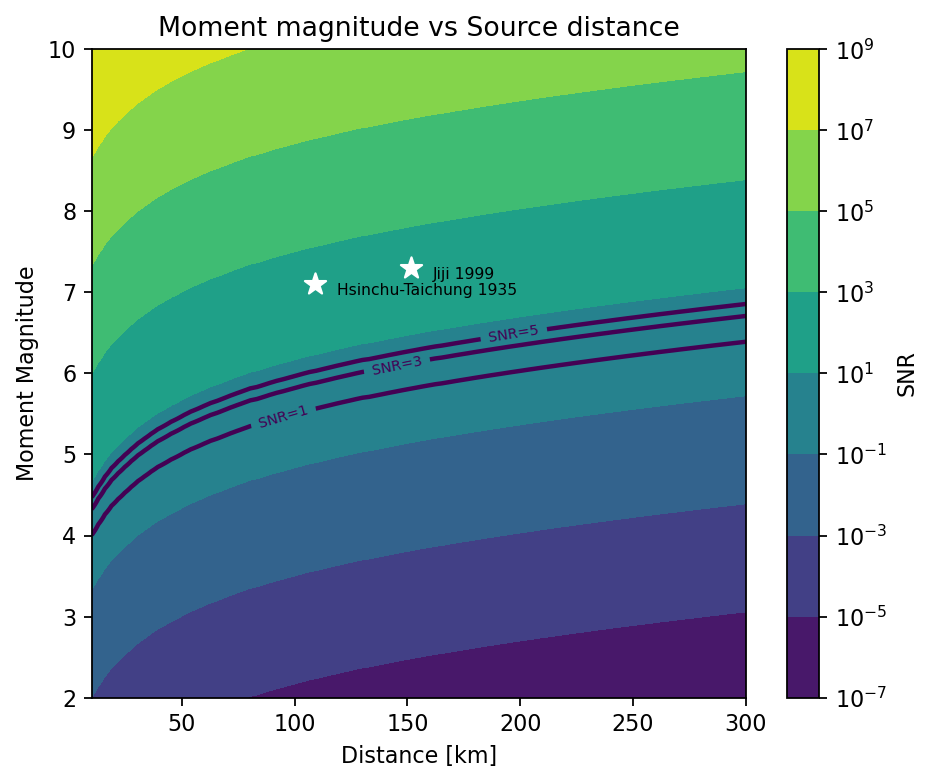}
\caption{Logarithmic SNR contour map as a function of earthquake moment magnitude and source--detector distance. The white stars indicate the locations of the 1999 Hsinchu--Taichung earthquake and the 1935 Jiji earthquake.}
\label{fig:EQ_d}
\end{figure}

As illustrative examples, two historically significant earthquakes are projected onto the distance--magnitude parameter space shown in Fig.~\ref{fig:EQ_d}: the 1999 Hsinchu--Taichung earthquake and the 1935 Jiji earthquake. Using the reported magnitudes and source distances from the CWA catalog \cite{CWA_EarthquakeCatalog}, both events lie above the SNR = 5 contour under the assumed detector model.

Although this comparison does not constitute a validation of the earthquake signal model, it provides a useful benchmark for interpreting the detectability estimates in terms of real seismic events. The locations of these earthquakes in the distance--magnitude plane indicate that the predicted detection region is populated by naturally occurring earthquakes rather than only by extreme or exceptionally nearby events.

More generally, the contour structure reflects the expected scaling of prompt gravity signals, with the SNR increasing for larger earthquake magnitudes and shorter source--detector distances. Together with the historical-event distribution shown in Fig.~\ref{fig:EQ_maphist}, these results suggest that earthquake-induced gravity perturbations could represent a realistic observational target for future CHRONOS measurements.

\subsection{Limitations and future work}

Several simplifying assumptions have been adopted in the present study. The Rayleigh-wave calculations assume a homogeneous medium and employ a phenomenological seismic quietness factor to characterize local conditions. Likewise, the atmospheric contribution is estimated using a simplified broken power-law model rather than site-specific infrasound measurements.

For earthquake signals, the adopted model is based on idealized seismic source descriptions and does not account for complex fault geometries, heterogeneous geological structures, or realistic wave propagation effects. These simplifications are appropriate for an initial feasibility study but introduce uncertainties in the predicted signal amplitudes.

Nevertheless, the principal conclusions of this work are expected to remain robust. In particular, the dominance of Rayleigh-wave-induced gravity-gradient backgrounds over atmospheric contributions, as well as the strong dependence on local site conditions, are consistent with previous studies of terrestrial Newtonian noise. Similarly, the predicted scaling of earthquake-induced gravity signals with source distance and moment magnitude follows directly from established gravitational and seismic considerations.

Future work should incorporate measured environmental data from candidate sites, realistic three-dimensional geological models, and more sophisticated earthquake source descriptions. Such improvements will enable more accurate estimates of both stochastic gravity-gradient backgrounds and transient earthquake-induced gravity signals, while providing a more rigorous assessment of their detectability with CHRONOS.

\section{Conclusion} \label{sec:conclusion}
In this work, we have investigated the gravity-gradient noise environment relevant to the proposed CHRONOS detector and evaluated its implications for both low-frequency gravitational-wave detection and transient geophysical signal observation. We developed a unified framework describing Rayleigh-wave seismic coupling, atmospheric infrasound coupling, and prompt gravitational perturbations from earthquake rupture processes, and projected these signals onto the torsion-bar detector response using a full mechanical–optical model.

Using representative site parameters, we find that Rayleigh-wave-induced gravity-gradient noise is the dominant environmental contribution in the sub-Hertz band and can exceed the projected CHRONOS sensitivity below approximately $0.5\,\mathrm{Hz}$. In contrast, atmospheric gravity-gradient noise remains several orders of magnitude smaller throughout the frequency range considered. The results further indicate that the overall Newtonian-noise level is governed primarily by the seismic quietness factor, while detector depth plays a comparatively weaker role.

We also evaluated the detectability of prompt gravitational signals generated by earthquakes. For a representative $M_w = 5.2$ event, we find that earthquakes within $\sim 90$ km of the detector may be detectable, while a $40$ km event yields $\mathrm{SNR} \approx 3.62$ integrated over the CHRONOS sensitivity band. The corresponding signal becomes comparable to the detector noise floor in the $\sim 0.2$–$0.6,\mathrm{Hz}$ region, and arrives ahead of seismic P-waves by $\sim 5$ s (ranging from $2.9$ to $6.9$ s depending on assumed P-wave velocity).

Historical earthquake catalogs further suggest that several potentially detectable events may occur each year at the proposed site, indicating that prompt gravity signals from earthquakes may be observable on realistic timescales.

Overall, CHRONOS provides a promising testbed for sub-Hertz gravity-gradient physics.

While Newtonian noise is traditionally regarded as a limiting noise source for low-frequency gravitational-wave detectors, our results demonstrate that it also constitutes a measurable geophysical signal. CHRONOS therefore provides a unique opportunity to study terrestrial gravity fluctuations while simultaneously advancing sub-Hertz gravitational-wave detector technology.

Future work will refine these estimates using site-specific seismic correlation models and improved atmospheric spectra.
\begin{acknowledgments}
We would like to express our sincere gratitude to W.Garcia, M.Hazumi, Y-C.Lin, and M.F.Vega for their valuable discussions and continuous support throughout this work. We also acknowledge the support and collaborative environment provided by the Department of Physics and the Center for High Energy and High Field (CHiP) at National Central University, the Institute of Physics, Academia Sinica, the National Institute of Physics, University of the Philippines Diliman. Y.I. and D.T. are supported by the National Science and Technology Council (NSTC) of Taiwan under Grant No. 114-2112-M-008-006, and by Academia Sinica under Grant No. AS-TP-112-M01.
\end{acknowledgments}

\section*{Conflict of Interest Statement}
The authors declare that they have no known competing financial
interests or personal relationships that could have appeared to
influence the work reported in this paper.

\section*{Data Access Statement}
The data and computational methods used in this study are available from the corresponding author upon reasonable request. All results presented in this work can be reproduced using the procedures and parameters described in the manuscript.

\section*{Ethics Statement}
This research did not involve human participants, human tissue, animals, or personal data. Therefore, no ethical approval was required for this study.

\bibliographystyle{apsrev4-2}
\bibliography{bibfile}

\end{document}